# Ludvig Lorenz (1867) on Light and Electricity

Helge Kragh*

**Abstract**: Independent of Maxwell, in 1867 the Danish physicist L. V. Lorenz proposed a theory in which he identified light with electrical oscillations propagating in a very poor conductor. Lorenz's electrodynamic theory of light, which formally was equivalent to Maxwell's theory but physically quite different from it, was published in well-known journals in German and English but soon fell into oblivion. In 1867 Lorenz also published a paper on his new theory in a semi-popular Danish journal which has generally been overlooked. This other paper is here translated into English and provided with the necessary annotations.

**Introduction**

The name of the Danish physicist Ludvig Valentin Lorenz (1829-1891), a lecturer at the Royal Military High School in Copenhagen, is presently associated with four different areas of physics.[1] He found in 1869 the relationship between the refractive index of a transparent substance and its density known as the "Lorenz-Lorentz law." In 1881 he extended the Wiedemann-Franz law on the ratio of a metal's electrical conductivity and its thermal conductivity to include also the temperature dependence. The names "Wiedemann-Franz-Lorenz law" and "Lorenz number" derive from this work. In his last contribution to physics, dating from 1890, Lorenz developed on a non-electromagnetic basis a comprehensive theory of the scattering of light by a sphere. The theory was largely equivalent to the later theory of Gustav Mie dating from 1908. Hence the term "Lorenz-Mie theory" entered the physicists' vocabulary.

   Moreover, his name is also known from the "Lorenz gauge" in electrodynamics which goes back to an electrical theory of light he presented in 1867 and in which he introduced the concept of retarded electromagnetic potentials. In sharp contrast to Maxwell's electromagnetic theory of light published two years earlier, Lorenz dismissed the ether as a superfluous and methodologically flawed concept. Although his paper failed to make an impact on contemporary physics, it was critically discussed by Maxwell and several years later it attracted the interest of important physicists such as George FitzGerald, Oliver Lodge and Pierre Duhem. More recently aspects of Lorenz's non-Maxwellian and non-ether theory have been reviewed by a number of physicists and historians of physics.[2] There were several reasons for the theory's peripheral role in late nineteenth-century electrodynamics, one of them being that Lorenz never followed up on his 1867 theory which remained isolated from his later works and also, of course, from Maxwell's field theory.

---

* Niels Bohr Institute, University of Copenhagen. E-mail: helge.kragh@nbi.ku.dk.
[1] For Lorenz and his works in physics, including references to the primary literature, see Pihl (1972), Kragh (1991) and Keller (2002). A detailed English biography of Lorenz is under preparation and is expected to be published later this year.

      Lorenz presented his theory to the Royal Danish Academy of Sciences and Letters at a meeting of 25 January 1867 and later in the year it was published in Danish in the transactions (*Skrifter*) of the Academy. A German translation appeared in the widely read *Annalen der Physik und Chemie* which was again translated into English in *Philosophical Magazine* under the title "On the Identity of the Vibrations of Light with Electrical Currents."[3] Moreover, the paper was carefully reviewed in the leading abstract journal *Fortschritte der Physik*.[4] Lorenz's electrical theory of light was thus widely circulated and known to the international physics community. A French version only appeared in 1898 when the first volume of *Oeuvres Scientifiques de L. Lorenz* was published.[5]

      It is less well known that in 1867 Lorenz also published a popular exposition of his theory which adds some facets not to be found in his scientific paper.[6] The Danish paper "On Light" appeared in the semi-popular journal *Tidsskrift for Physik og Chemi* (Journal of Physics and Chemistry) founded in 1862. Although the paper is purely qualitative and without equations it is of some interest as it complements the technical paper and includes passages throwing light on Lorenz's methodology and his ideas of optics and electricity. For this reason the paper is a historical resource that should be available also to the large majority of scientists and historians unable to read Danish. An annotated translation follows.

---

[2] Rosenfeld (1956); Whittaker (1958), pp. 267-270; Kaiser (1981), pp. 157-162; Darrigol (2000), pp. 212-213; Jackson and Okun (2001); Keller (2002); McDonald (2016).
[3] Lorenz (1867a); Lorenz (1867b).
[4] Radicke (1870).
[5] Lorenz (1898-1904), vol. 1, pp. 173-196.
[6] Lorenz (1867c).

## On Light

What is light? To get a closer insight into this natural agency will surely be of interest to everyone with a sense for its grand meaning; and also because it is the most important intermediator between us and the outer world and the only messenger from the distant world globes, indeed for everything living and moving in nature. If the question concerns only the path of light and its swift motion through space and bodies, and how to find the laws for the motion, then science has a perfectly satisfying answer. In this case, where the question can be approached mathematically, it turns out that the phenomena of light are in close agreement with the exact and ideal results derived from mathematical thinking. But if we want to know more about the true nature of light – the very foundation of its actions – we have to admit that we are far from an adequate answer. The reason is especially that we know very little or even nothing of what happens in the deep interior of the bodies which is inaccessible to our senses and yet is the ultimate source of all physical action.

It was once assumed that light consist of tiny particles expelled from luminous bodies, an assumption which still in our century was defended with great skill. However, as more became known about the properties of light one was forced to assume that *light is a motion of waves*, meaning that the action of light consists in periodic motions or *vibrations propagated in forward direction*. Thus, it is not something of a material nature emitted from the luminous body but a motion in propagation. To this picture was associated a *medium* for the motion, for there cannot be empty space where a ray of light passes. This theory must be considered perfectly scientifically justified and for this reason be irrefutable at any time. Moreover, we can determine with great accuracy the enormous speed of the motion and the number of oscillations in any given time; and that notwithstanding that one needs millions of years to count the number of oscillations that a ray of light transmits to the eye in a second. And yet, if we proceed with the analogy between light and other forms of wave motion the results become more dubious. People have imagined that the medium for the motion of light was a special substance which combined extraordinary tension with imperceptible weight; that light was oscillations in this "ether" in analogy to sound consisting of oscillations in the material bodies. But this idea became increasingly doubtful, for light vibrations are different. They do not move in the direction of the light ray, as one might expect, but only perpendicular to it.

In an earlier paper in this journal's first volume[7] I accounted for my view concerning the theory of light. I have shown that one must disregard all physical hypotheses about the nature of light and base its laws solely on facts. Having completed this part of the theory of light (in Pogg. Ann., vol. 121),[8] the next step must be to establish a connection between these laws and the laws for other forces. The endeavours to search for connections between the various forces have been a significant reason for the progress of recent science; the idea that the various forces in nature are merely different manifestations of the one and same force has proved itself more fertile than all physical theories. It turned out that only one further step along the already established road had to be made, and this step leads to the remarkable result that *the vibrations of light are electrical currents*.

---

[7] Lorenz (1862).

To prove this one must know how the electrical currents propagate through bodies. In any point the current depends on the electricity of the surrounding parts of bodies, whether it is in motion as an electrical current or is at rest as static electricity. In the latter case it produces by "polarization" a separation of the two kinds of electricity in all the points of the body, and the electrical currents act in the same way if they decline or increase in strength. By polarization and "induction" the two kinds of electricity are separated and an electrical current is generated. After the corresponding laws had been established, Kirchhoff could finally present it all in a mathematical form and from his equations the problem could be solved.[9]

Regarding the plausibility of these equations it should be noticed, on the one hand, that they could be considered the correct expression for experimental results in so far that they match the accuracy of the experiments; on the other hand, they lack the theoretical foundation without which they would not be the exact expression for the law. To understand this, consider a body which in a point A suddenly receives some amount of electricity. As a result, in other points of the body a separation of the two kinds of electricity will be produced by polarization. The corresponding "electromotive" force will in part be proportional to the amount of electricity received in the point A and in part be inversely proportional to the square of the distance from this point. The law is exact in so far that it is simple and in agreement with the laws for other forces acting at a distance.

But the question is if the time for this [electrical] action is precisely determined. Does the action occur instantaneously in all parts of the body whether they are close to or far from the source point? Or does it take some time for the action to propagate so that it first arrives to the closer parts of the body and only later on to the points more far away? This question is closely connected to another one, namely if these forces really act at a distance or propagate from point to point through the intervening space such as suggested by Faraday. If the latter is the case it must take some time however small for the electrical actions to be transmitted from one place to another. If the speed of propagation is very high, say of the order of that of light, we would be unable to detect it experimentally and yet it does not follow that we can ignore it. As soon as we formulate this either-or question[10] the choice leaves no doubt, for it is a choice between the particular and the general encompassing the particular. In such a situation one must of course choose the general. In other words, one must assume that the electrical action does not propagate instantaneously but *propagates with a certain velocity*; then one can always return to the particular case by assuming the velocity to be infinite or the action at any distance to occur instantaneously. As far as experiments are concerned they merely show that the velocity must be very great but so far undetermined. Experiments can never prove that it is infinite and that the action appears instantaneously. But really, this is the assumption behind the laws that Kirchhoff and others before him have stated for the electrical actions and it can only be justified by a much deeper knowledge of the nature of electricity than we possess at present.

---

[8] Lorenz (1864). "Pogg. Ann." was a common abbreviation of *Annalen der Physik und Chemie*, Europe' foremost scientific journal edited by the German physicist Johann C. Poggendorff (1824-1876).

[9] Kirchhoff (1857).

To assume that Kirchhoff's formulae are incomplete and yet approximately correct is the same as regarding them as *the first term in a series expansion*. They seem indeed to have the character of such a series expansion. I have assumed that the electrical actions need a very short time to propagate and tried to generalise the formulae accordingly. In this way I have arrived at different formulae which are somewhat simpler than Kirchhoff's and, when expanded in a series, coincide with Kirchhoff's in their first term. The next terms in the series include a very small quantity raised in increasing powers and they are insignificant in ordinary experiments with electrical currents; they only become of some importance when they current change significantly in an extremely small time interval. In this modified form the equations show that in poor conductors electricity can exist as periodic currents vibrating in both directions but propagating in none of these directions as they propagate in a direction perpendicular to them. What turns out to be *possible* in calculations deduced from really existing laws and conditions will always turn out to correspond to reality.[11] Where, then, should we look for these periodic currents, propagating more easily the more poorly they are conducted in the body and only in a direction perpendicular to the current, if not in *the ray of light*? After all, the vibrations of light are periodical and perpendicular to the direction of light; moreover, light can only pass through extremely poor conductors.

It turns out that the velocity of propagation in poor conductors is just the same as the velocity of light and that it is possible to determine the velocity of light in air solely from

---

[10] This is possibly an allusion to Søren Kierkegaard's "either-or" philosophy as expounded in his *Philosophiske Smuler* (Philosophical Fragments) from 1844. In his youth Lorenz studied Kierkegaard's work which influenced his thinking.

[11] The statement is a summary declaration of the so-called principle of plenitude. According to this metaphysical principle, what is allowed according to the laws of nature and hence can potentially exist also has a real existence. Whether expressed explicitly or just used implicitly the principle has played an important heuristic role in a wide range of sciences, including chemistry, astrophysics, cosmology and modern elementary particle physics.

Weber's experiment with an accuracy no less than the one determined in different ways.[12] Moreover, it turns out that the equations for the electrical currents can be transformed in such a way that they, apart from one term, *agree completely with the equations I have previously found for the vibrations of light* and from which the entire theory of light can be deduced.[13] The disagreement due to the mentioned term merely serves to confirm the correctness of the theory since the term becomes significant only for good conductors such as the metals; it shows that they must absorb light in accordance with experience, whereas the term disappears for very poor conductors. From this we infer that even the least degree of transparency indicates that the body is a very poor conductor compared to the metals. On the other hand, the reverse is not necessarily the case, for opaqueness can be due to other causes of which lack of homogeneity is one example.

The theory thus cast a new light on several facts such as the poor electrical conductivity of transparent bodies, the opaqueness of good electrical conductors, and a certain agreement between the velocity of propagation of light and that of electricity; these facts suggest a connection between light and electricity which has long been suspected but an explanation of which has been missing. It would be extremely difficult to demonstrate by means of direct experiments that light vibrations are electrical currents, the reason being that the electrical currents in a ray of light vibrate billions of times per second. And yet there are known facts indicating that vibrations of light might be transformed to electrical currents, namely when light hits upon the interface between two different metals. The interface acts like a kind of electrical valve favouring the propagation in one direction and not the other.[14] Still, the real causes are not revealed by these facts, for they depend on the molecular constitution of the bodies.

Having demonstrated that oscillations of light are electrical currents we could further ask, what is an electrical current? On the assumption that an oscillation of light is the same as oscillating parts of the ether, the electrical current is nothing but a progressive motion of the ether – a real and material current of some sort of liquid. But although this is an often-held opinion it is completely untenable. Sure, the assumption may lead to the correct equations for the oscillations of light, but only by regarding them as infinitesimally small displacements of ethereal parts. As soon as we deal with the finite displacements constituting the electrical currents the equations will appear in quite a different form. The identity of the equations for electrical currents and those of light vibrations demonstrates that we always have to do with small relative motions, both when the electrical currents are vibrating back and forth and when they are propagating forwards. Neither in the case of the electrical current nor in that of the ray of light is there anything of a material kind which moves. We are dealing only with propagating *molecular motions*.[15] What these motions are, more exactly, is a question we cannot yet answer with any degree of certainty. I once proposed the hypothesis that the oscillations of light were rotating oscillations of material parts and that the direction of the oscillations is along the axis of rotation.[16]

---

[12] The reference to the experiment of Wilhelm Weber (1804-1891) is to a famous series of experiments made by Weber and Rudolf Kohlrausch (1809-1858). For an English translation of the Weber-Kohlrausch paper from 1856, see Assis (2003).

[13] The reference is to Lorenz (1864).

[14] Lorenz might have thought of the photovoltaic effect discovered by A. E. Becquerel (1820-1891) in 1839 or possibly of the thermoelectric Seebeck effect named after the German physicist Thomas Seebeck (1770-1831) who discovered or rediscovered it in 1821.

On this hypothesis one must consider the electrical current to be a continual rotation of the material parts and the direction of the axis of rotation becomes the direction of the current. But other kinds of molecular motion are conceivable and it is unlikely that the problem can be easily solved. Besides, to add to the assumption of molecular motion the assumption of an ether would be unreasonable; because, it is a new non-substantial medium which has been thought of only because light was conceived in the same manner as sound and it hence had to be a medium of exceedingly large elasticity and small density in order to explain the large velocity of light. However, the hypothesis is superfluous if the velocity does not depend on elastic forces but on quite different molecular forces. What we know about the huge power of the molecular forces makes it understandable that they are able to generate motions with such a great velocity of propagation. If we need to assume some medium for light between the celestial globes, we do not have to conceive it as different from the known gases.[17] On the whole it is most unscientific to fabricate a new substance when its existence is not revealed in a much more definite way.

Among the certain results of this investigation is that it takes time for the actions of electricity to propagate from one place to another. What Rømer taught us about light 200 years ago is valid also for the electrical forces.[18] We might well add that it is valid also for other forces such as the gravitational attraction and assume that in general no action can propagate instantaneously and thus be all over in space at the same time.[19] The basic assumption is that the forces or at least the electrical forces propagate successively from one point to another in the bodies. They appear to act at a distance, but in reality they do not reach farther than to the nearest surrounding molecules.

---

[15] Darrigol (2000), p. 213 points out that Lorenz "identified the optical ether with a bad conductor." Despite his rhetorical dismissal of the ether he needed a conducting medium.

[16] For Lorenz's earlier allusion to light being rotational, which was also an idea vaguely entertained by his former teacher H. C. Ørsted, see Lorenz (1863).

[17] Conceptions of the ether as a rarefied gas were known at the time and later in the century. A few chemists suggested that the ether consisted of a new gaseous element lighter than hydrogen. See Kragh (1989).

[18] In 1676 Lorenz's compatriot Ole Rømer (1644-1710) proved by astronomical measurements that the velocity of light is finite ($c \sim 227{,}000$ km/s).

[19] The suggestion that gravity propagates with of a finite velocity was not mentioned in Lorenz (1867a).

-oOo-